\documentclass{aa}
 
\usepackage{graphics,times}

\begin{document}

\thesaurus{04         
          (08.12.1;           
           08.12.2;           
           08.12.3;           
           08.23.1;           
           10.19.1            
           10.19.2)}          

\title{The luminosity function of white dwarfs and M dwarfs \thanks{Based on observations collected at the European Southern Observatory (ESO), La Silla, Chile and at the Nordic Optical Telescope (NOT), La Palma}}

\subtitle{using dark nebulae as opaque outer screens}

\author{L. Festin}

\offprints{L. Festin}

\institute{Astronomical Observatory,
	      Box 515,
	      S-751 20 UPPSALA,
              Internet: leif@astro.uu.se
             }

\date{Received            ; accepted           }

\maketitle

\begin{abstract} 
By using dark nebulae as opaque outer screens, the luminosity function of white dwarfs and M dwarfs has been studied. High-extinction areas towards the Orion A, Serpens and $\rho$ Oph cloud complexes were observed, covering a volume corresponding to 464 pc$^{3}$ in the solar neighbourhood, complete to $M_\mathrm{V}\sim16.5$. Foreground stars were selected by $VRI$ photometry and photometric parallaxes.

The resulting foreground sample consists of 21 M dwarfs and 7 white dwarfs. The derived M-dwarf luminosity function is consistent with previous studies, showing no substantial upturn beyond $M_\mathrm{V}=16$. The 7 white dwarfs imply a local density of 0.013 ($\pm0.005$)\,M$_{\sun}$\,pc$^{-3}$ ($\sim\,$15\,\% of the dynamical mass in the solar neighbourhood) supporting other recent results but obtained with a completely different method.

For the clouds, foreground-star data were used to derive independent cloud distances, while the background stars and a simple model of the Milky Way gave reliable cloud extinctions.


\end{abstract}

\section{Introduction}
The dark matter affects astronomy on all distance scales,  single galaxies, groups of galaxies and the Universe as a whole. 
The dynamical mass in the solar vicinity has recently been determined from Hipparcos data to $0.076\,\pm\,0.015$ M$_{\sun}$\,pc$^{-3}$ (Cr\'ez\'e et al. \cite{creze97}), leaving little space for local dark matter, since the visible matter in form of stellar-like objects and the interstellar medium comprise $\sim\,0.08$ M$_{\sun}\,$pc$^{-3}$. 

The stellar contribution to the local mass density derived by Jahreiss \& Wielen (\cite{jahreiss97}) is 0.039 M$_{\sun}$\,pc$^{-3}$, which agrees within errors with most other recent figures (e.g. $0.042\,\pm\,0.008$ in M\'era et al. (\cite{mera96}), $0.05 \pm 0.01$ in Kroupa et al. (\cite{kroupa93}) and $0.034$ in Gould et al. (\cite{gould96})).
Similar amounts of matter have been suggested to reside in the interstellar medium, $\sim\,0.04$ M$_{\sun}$\,pc$^{-3}$ (Haywood et al. \cite{haywood97};  Cr\'ez\'e et al. \cite{creze97}; Bahcall et al. \cite{bahcall92} and references therein). However, present estimates of the local surface density of the interstellar medium range from 6 to 13 M$_{\sun}$\,pc$^{-2}$, indicating a substantial uncertainty also in the space density.

The photometric surveys by Gould et al. (\cite{gould96}) and Kirkpatrick et al. (\cite{kirkpatrick94}) give 0.011 M$_{\sun}$\,pc$^{-3}$ for M dwarfs in the mass interval 0.1--0.5 M$_{\sun}$. For the nearby stars the corresponding numbers are 0.013 and 0.015 M$_{\sun}$\,pc$^{-3}$ respectively for Reid \& Gizis (\cite{reid97}) (hereafter RG97) and Jahreiss \& Wielen (\cite{jahreiss97}). The discrepancy between the photometric and the nearby samples is significant and has been suggested to arise from unresolved binarity (Kroupa \cite{kroupa95}) and structure in the $M_\mathrm{V}$ vs $V-I$ relation that has not been taken into account (RG97).
For the white dwarf (WD) mass density, the 0.003 WDs\,pc$^{-3}$ ($\sim$\,0.002 M$_{\sun}$\,pc$^{-3}$) given in Liebert et al. (\cite{liebert88}) (hereafter LDM88), is less than 3\,\% of the local dynamical mass. 
There are, however, recent investigations that point to a much higher mass hidden in WDs. Ruiz \& Takamiya (\cite{ruiz95a}) (hereafter RT95) derived, from a large proper motion survey, a factor three higher space density than in LDM88. A similar value was obtained by Oswalt et al. (\cite{oswalt96}) from WDs in binaries. This revision of the WD space density (supported also by this paper) indicates that as much as 15\,\% of the local dynamical mass can be hidden in WDs.

Thus, although the bulk of the local mass seems to be identified, the role of the late M dwarfs and WDs in this context is still not clear and remains a question to discuss.

In this paper the subject is addressed in a survey of opaque nebulae, obtaining volume limited and dynamically unbiassed samples of foreground M dwarfs and WDs. This method was first proposed by Herbst \& Dickman (\cite{herbst83}) and has previously been used by Jarrett (\cite{jarrett92},\cite{jarrett95}) and Jarrett et al. (\cite{jarrett94}) (hereafter JDH94), whose results will be discussed in Sect.~\ref{sec-jarrett}. The present survey covers a volume corresponding to 464 pc$^{-3}$ in the solar neighbourhood and is complete to $M_\mathrm{V} \sim 16.5$. The completeness limit is 2 magnitudes deeper than the previous surveys by Jarrett and the volume twice as large.

In \S\,2 the selection of the target clouds is described. \S\,3 outlines the observations and reduction procedures. The extinctions of the clouds are derived in \S\,4. The selection of foreground stars, including colour-colour analysis and photometric parallaxes is decribed in \S\,5. Constraints on our foreground sample obtained from other sources are given in \S\,6. The M-dwarf luminosity function (LF) and WD space densities, including corrections for multiplicity and galactic density gradients are derived in \S\,7. In \S\,8 the results of this paper are compared to other recent findings, with special emphasis on the dark cloud survey described in JDH94. The conclusions and some ideas for the future are given in \S\,9.

\section{Selecting target clouds}
A suitable target cloud should fulfill certain criteria. 
Its distance has to be known in order to match integration times to the faintest stars at the cloud's edge. It has to be sufficiently opaque ($A_\mathrm{V}\ga 5$ mag) to screen out the vast majority of background stars and galaxies. The high extinction areas must be large enough to provide a reasonable number of foreground stars. It is also desirable to avoid the immediate neighbourhood of star-forming regions, since young overluminous stars within, or just in front of, the cloud may be mistaken as nearby older field M dwarfs.

The final selection includes the following cloud complexes, Orion A (southern part of L1641, 480 pc, $l=213\degr$, $b=-19\degr$), Serpens ($\sim\,4\degr$ south from the Serpens cloud core, 200--250 pc, $l=27\degr$, $b=3\degr$) and two separate areas in $\rho$ Oph (L1688 and L1689, 160 pc, $l=354\degr$, $b=16\degr$).  Cloud distances for the original selection were taken from Hilton \& Lahulla (\cite{hilton95}).

\subsection{Cloud distances}
The photometric distances and the total number of M dwarfs in front of Orion A indicate a distance of 400 rather than 480 pc. This result is consistent with the 390 pc in Anthony--Twarog (\cite{anthonytwarog82}), derived from a subsample in the Warren \& Hesser (\cite{warren78}) catalogue of stars associated with the Orion complex. In order to confirm this finding, the Hipparcos catalogue (ESA \cite{hipparcos}) was searched, resulting in parallaxes with standard errors less than 1 mas for 92 of the Warren stars, showing a well-defined peak at $\sim\,400$ pc.

For Serpens, Straizys et al. (\cite{straizys96}) gave a mean distance of 259 pc for 18 stars associated to the cloud. Their total sample consists of 99 stars, foreground, embedded and background. The low-extinction part of their sample indicates an extinction wall at $\sim$\,255 pc. Of the 99 stars, 13 were measured by Hipparcos, revealing a few cases of misclassified luminosity classes, but basically confirming a distance of $\sim$\,260 pc.

Eleven of the stars that were assigned a high probability to be associated with the $\rho$ Oph complex by Elias (\cite{elias78}) were identified in the Hipparcos catalogue and show a sharp peak at 150 pc. 

The distances finally adopted in this paper are 400, 255 and 150 pc respectively to the Orion A, Serpens and $\rho$ Oph complexes. Note that these clouds are extended irregular objects far from being sheet-like or spherical. The distances above were derived from observations of optically visible stars, and therefore mainly probe the front edge of the cloud, which though is sufficient for the purpose of this paper. The errors in the cloud distances are expected to be on the order of 10\,\%, giving a corresponding volume uncertainty of $\sim$\,30\%. 

\section{Observations and reductions}
The main body of data consists of $VRI$ photometry and was collected at the ESO NTT 3.55\,m telescope at La Silla, Chile. The Serpens and $\rho$ Oph areas were observed in Jun28--Jul01, 1995 and cover 536 and 816 arcmin$^{2}$ respectively. The Orion A field, observed in Jan24--26 1996, covers 158 arcmin$^{2}$. The same instrumental setup was used in both runs, EMMI RED with CCD \#36, TEK2048, each pixel corresponding to 0.27\arcsec. The unvignetted  field of view was  8.6' x 9.15'. 
Bias and flatfield corrections were done in a standard way with IRAF \footnote{IRAF (Image Reduction and Analysis Facility) is distributed by National Optical Astronomy Observatories (NOAO), which is operated by the Association of Universities for Research in Astronomy, Inc., under contract with the National Science Foundation.}.

The transformations to the Johnson $V$ and Kron-Cousins $RI$ systems were based on standard stars selected from Landolt (\cite{landolt92}). Since the observations were aimed at M dwarfs in the first place, care was taken to include the reddest dwarf stars in this list, G45-20 ($V-I=4.00$), G156-31 ($V-I=3.68$), G12-43 ($V-I=3.48$), G10-50 ($V-I=2.97$)  and G44-40 ($V-I=2.79$).  The transformation to the standard system includes corrections for extinction, zero-point offsets and linear colour terms. The rms of the standard star residuals across the whole fitted colour range ($0<V-I<4$) is 0.02 mag in $VR$ and 0.03 mag in $I$.

\subsection{Complementary NOT observations}
For Orion A, part of the $VRI$ photometric candidates were measured in $J$ using ARNICA, a NICMOS3 near-IR array at the Nordic Optical Telescope (NOT), La Palma, in Sep 1996. The $J$ calibrations were made via standard stars provided by the ARNICA team (Hunt et al. \cite{hunt95}). The Hunt transformation was adopted: $J_\mathrm{CIT}=J_\mathrm{ARNICA}+$ zero-point offset. The final $1\,\sigma$ error in the calibration of the $J$ magnitudes is 0.07 mag.

Low-resolution spectra of the two brightest WD candidates were obtained at the NOT in Nov 1997, using the ALFOSC, a low-resolution spectrograph/focal reducer combination instrument. Standard reductions were carried out with IRAF. It was not possible to apply a meaningful flux calibration, since the targets were always at airmasses between 2 and 3.5, leading to a substantial loss of light in the blue end of the spectra.

Since this survey was designed for detecting M dwarfs, the original $I$ exposure times were set rather short. The discovered WDs are bluer than the M dwarfs, resulting in uncertain  $I$ magnitudes in the original data. The WD $I$ photometry was improved by ALFOSC observations at the NOT in Nov 1997. Again standard methods were used in the reductions.

All instrumental magnitudes were extracted by an empirical growth-curve method outlined in Festin (\cite{festin97a}).\\ 

The $M_\mathrm{V}$ completeness limit of the survey was set by the $V$ data, as the magnitude at which the mean of the internal magnitude errors in a field equals 0.12 mag. This limit was justified by a star-count analysis in a previous study of the Pleiades (Festin \cite{festin98a}). Table~\ref{complimits} gives completeness limits and positions for all the observed fields.

\begin{table*}
  \caption[]{Observed areas and completeness limits. The photometric system is $RI$ Kron-Cousins (KC) and $V$ Johnson (J)}
  \begin{flushleft}
    \begin{tabular}{lrrrrccccc}
      \hline
      field id&\multicolumn{4}{c}{field centres}&&\multicolumn{4}{c}{completeness limits} \\
      \cline{2-5}\cline{7-10}
      &RA (2000)&DEC (2000)&\makebox[0.5cm][l]{$l$}&\makebox[0.5cm][l]{$b$}&&$I_{\rm KC}$&$R_{\rm KC}$&$V_{\rm J}$&$M_\mathrm{V}$ \\
      \hline
       Serpens A         & 18:27:29.1 &  -3:42:21 &   26.8931 &   3.6223 && 22.1 & 23.2 & 24.2 & 17.2  \\	    
       Serpens B         & 18:28:04.4 &  -3:42:21 &   26.9610 &   3.4921 && 22.0 & 23.3 & 24.4 & 17.4  \\	
       Serpens C         & 18:28:03.3 &  -3:49:57 &   26.8463 &   3.4377 && 20.4 & 22.3 & 23.9 & 16.9  \\	
       Serpens F         & 18:27:52.4 &  -3:34:16 &   27.0576 &   3.5986 && 21.3 & 22.5 & 23.7 & 16.7  \\	
       Serpens J         & 18:28:23.2 &  -3:34:18 &   27.1164 &   3.4847 && 21.3 & 22.9 & 23.4 & 16.4  \\	
       Serpens K         & 18:28:39.4 &  -3:45:28 &   26.9821 &   3.3390 && 21.5 & 22.9 & 24.3 & 17.3  \\	
       Serpens L         & 18:29:13.7 &  -3:45:28 &   27.0480 &   3.2125 && 22.2 & 23.6 & 24.5 & 17.5  \\	
       $\rho$ Oph E E    & 16:31:56.7 & -24:27:53 &  353.8780 &  15.9224 && 21.5 & 22.6 & 23.5 & 17.6  \\	
       $\rho$ Oph E F    & 16:32:49.3 & -24:29:24 &  353.9928 &  15.7544 && 21.4 & 22.7 & 23.7 & 17.8  \\	
       $\rho$ Oph E G    & 16:32:48.2 & -24:38:01 &  353.8765 &  15.6643 && 21.5 & 22.6 & 23.5 & 17.6  \\	
       $\rho$ Oph E H    & 16:33:22.5 & -24:40:14 &  353.9349 &  15.5415 && 21.5 & 22.7 & 23.7 & 17.8  \\	
       $\rho$ Oph E I    & 16:33:24.6 & -24:32:29 &  354.0423 &  15.6192 && 21.5 & 22.5 & 23.6 & 17.7  \\	
       $\rho$ Oph E J    & 16:33:23.3 & -24:25:16 &  354.1341 &  15.7009 && 21.4 & 22.6 & 23.6 & 17.7  \\	
       $\rho$ Oph W B    & 16:28:00.1 & -24:36:09 &  353.1580 &  16.5100 && 21.4 & 22.4 & 23.5 & 17.6  \\
       $\rho$ Oph W D    & 16:27:42.4 & -24:43:33 &  353.0152 &  16.4791 && 20.3 & 21.0 & 22.4 & 16.5  \\
       $\rho$ Oph W D'   & 16:27:42.3 & -24:28:45 &  353.2084 &  16.6420 && 21.3 & 22.5 & 23.4 & 17.6  \\
       $\rho$ Oph W E    & 16:27:27.7 & -24:36:09 &  353.0736 &  16.6023 && 20.9 & 22.6 & 23.5 & 17.7  \\
       $\rho$ Oph W G    & 16:27:09.1 & -24:27:49 &  353.1339 &  16.7469 && 20.0 & 21.5 & 22.6 & 16.7  \\
       Orion A F1        &  5:41:22.6 &  -8:34:35 &  212.7511 & -19.4421 && 21.5 & 23.4 & 24.5 & 16.5  \\
       Orion A F2        &  5:41:22.6 &  -8:25:57 &  212.6138 & -19.3793 && 21.5 & 23.4 & 24.5 & 16.5  \\
      \hline
    \end{tabular}
  \end{flushleft}
  \label{complimits}
\end{table*}

\section{Star counts and cloud extinctions}
The extinction offsets relative to a reference field were derived for each field by the method of star counts (Dickman \cite{dickman78}).  Since there are more background stars present in the $I$ fields than in $V$ or $R$, the $I$ band was used for the star counts. The derived extinction in $I$ is transformed to $V$ by the relation in Winkler (\cite{winkler97}), $A_\mathrm{V}=1.69A_\mathrm{I}$.

\begin{figure}
\resizebox{\hsize}{!}{\includegraphics{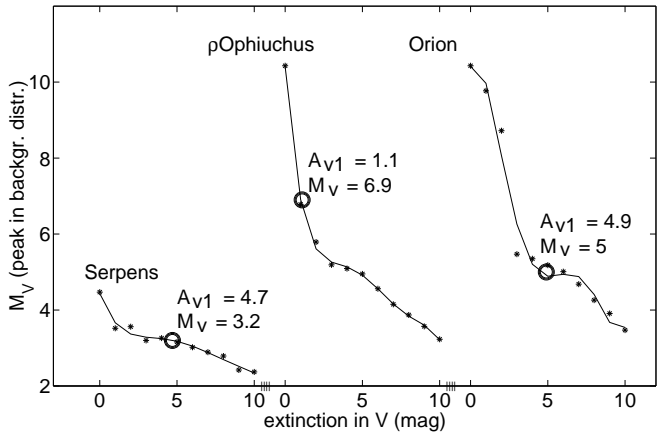}}
\caption{$M_\mathrm{V}$ at the peak in the through-shining background dwarf distribution as a function of the cloud's visual extinction. The solid line is a polynomial least-square fit to the calculated points (asterisks). The rings mark the three reference fields at the extinction $A_\mathrm{V1}$ as derived in Fig.~\ref{Figminext}}
\label{Figmagext}
\end{figure}

\begin{figure}
\resizebox{\hsize}{!}{\includegraphics{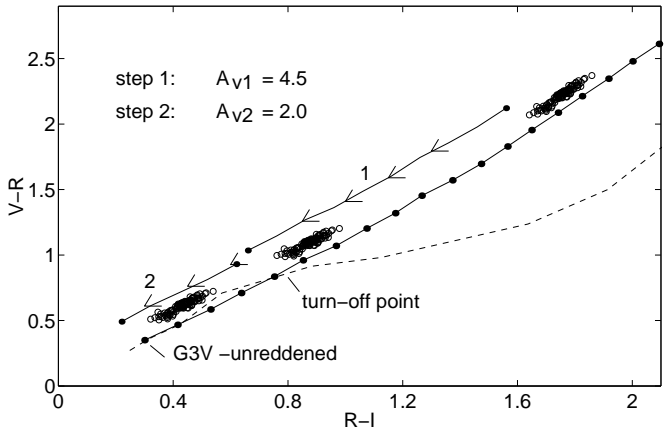}}
\caption{The two-step extinction estimation. The dashed line is the main sequence. The solid line is the G0V reddening curve in Jarrett (\protect\cite{jarrett92}). Each dot is a 0.5 mag step in $A_\mathrm{V}$. The arrowed solid lines show the dereddening steps. The rings demonstrate the location of a fictious set of stars}
\label{Figextoff}
\end{figure}

\begin{figure}
\resizebox{\hsize}{!}{\includegraphics{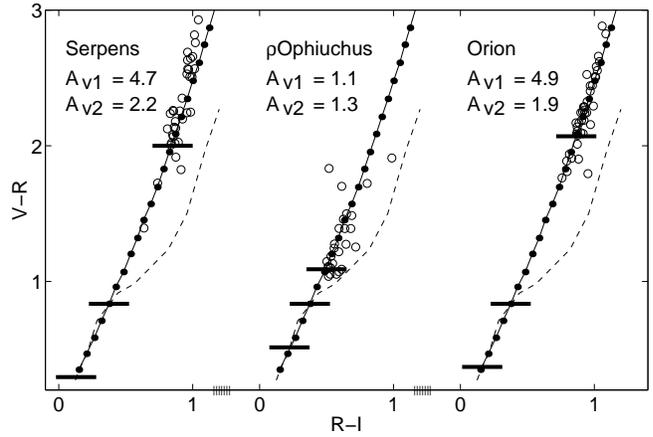}}
\caption{The extinction in the reference fields. The main sequence and reddening curves are as in Fig.~\ref{Figextoff}. The rings are real stars and the horizontal bars show steps 1 \& 2 of Fig.~\ref{Figextoff}}
\label{Figminext}
\end{figure}

In order to estimate absolute extinctions in the reference fields, a two-step method was used. A first estimate is obtained by shifting the blue edge of the bulk of the background stars along the reddening curve to the turn-off point (step 1 in Fig.~\ref{Figextoff}). Since the majority of the background stars are reddened beyond the selected blue edge and their intrinsic colours are significantly bluer than the turn-off point ($\sim$M0V), the corresponding extinction offset is a crude lower limit of the real value. 
Although this lower limit is sufficient for most cases, it can be further refined (step 2 in Fig.~\ref{Figextoff}) by using knowledge of the galactic structure and the LF. The Galaxy model of Bahcall \& Soneira (\cite{bahcall80}) was used together with the LFs in Gould et al. (\cite{gould96}) ($M_\mathrm{V}>8$) and Scalo (\cite{scalo86}) ($M_\mathrm{V}<8$) to derive the distribution of background stars as a function of cloud extinction. Figure~\ref{Figmagext} shows $M_\mathrm{V}$ at the peak in the distribution of through-shining stars as a function of $A_\mathrm{Vcloud}$. The observed fields are at a low galactic latitude and suffer from high extinction. The background stars are therefore dominated by the thin-disk population, which makes the single-component disk of Bahcall appropriate to use. Uncertainties inherent in the model would mainly arise from systematic errors in the scale height vs spectral type relation. Such effects are small enough to be dominated by the variation in the LF of the background stars. Errors in step 2 that may arise from uncertainties in the LF are expected to be small, since the visible background stars predominantly consist of dwarfs between F and early K, i.e. stars for which the LF is well defined. Steps 1 \& 2 are shown for the three reference fields in Fig.~\ref{Figminext}. Note that step 1 is a lower limit of the extinction derived independently from any assumptions of galactic structure or the LF. The error in step 2 is estimated to be $\sim$ 0.5 mag. The derived extinction of the reference field, $A_\mathrm{Vref}=A_\mathrm{V1}+A_\mathrm{V2}$, is still only a lower limit, since we may now use the better estimate $A_\mathrm{Vref}$ instead of $A_\mathrm{V1}$ in step 2, and iterate further. However, we restrict ourselves to two steps and regard the derived value to be a conservative lower limit with an error of $\sim$ 0.5 mag in $A_\mathrm{Vref}$.

\subsection{Orion A} 
The whole area has a high roughly constant extinction. The two-step procedure was applied to the entire field, resulting in $A_\mathrm{Vref}>6.8$. From $^{13}$CO data in Fukui \& Mizuno (\cite{fukui91}) an independent value of $A_\mathrm{V}\sim7$ was derived using the $A_\mathrm{V}-^{13}$CO relation in Bally et al. (\cite{bally91}).

\subsection{Serpens and $\rho$ Oph}
The fields were divided into overlapping squares of 1\arcmin\,x\,1\arcmin, separated by 15\arcsec. The number of stars within each square was summed in 0.1 mag bins, providing the extinction relative to the reference field.

The results for Serpens and $\rho$ Oph respectively are $A_\mathrm{Vref}>6.9$ and $A_\mathrm{Vref}>2.4$. The same reference field was used for both areas in $\rho$ Oph, since they are separated by only $1 \degr$.

The extinctions at the centres of the two areas in $\rho$ Oph were estimated from $^{13}$CO column densities in Loren (\cite{loren89}).  Using the same $A_\mathrm{V}$--$^{13}$CO relation as in the Orion case, $A_\mathrm{V}\sim 11$ was deduced. The present survey gives a lower limit of $A_\mathrm{V}=7.5$. For the  reference field, the $^{13}$CO gave $A_\mathrm{V}\sim 3$, whereas this survey gives a lower limit of $A_\mathrm{Vref}=2.4$.

For both Orion A and $\rho$ Oph, the lower limits from the present survey are in good agreement with independent values derived from $^{13}$CO data. This shows that reliable lower limits on the extinction can be obtained directly from the $VRI$ data. The Serpens extinction, derived using the same method as for Orion and $\rho$ Oph, is therefore likely to be a reliable lower limit as well.

\section{Selecting foreground stars}
\label{secextract}
To really make use of the clouds as screens, crowded regions must be avoided. A conservative upper limit on the average surface density of stars brighter than the completeness limits was set at 4 per arcmin$^{2}$. The remaining areas are for $\rho$ Oph 689 arcmin$^{2}$ and for Serpens 412 arcmin$^{2}$, 84 and 77 \% of the respective original fields. For Orion A the entire field (158 arcmin$^{2}$) is sufficiently sparse.

Figures of the target areas, including selected foreground stars and excluded regions are shown in App. A.

\subsection{Colour-colour analysis}
\begin{figure}
\resizebox{\hsize}{!}{\includegraphics{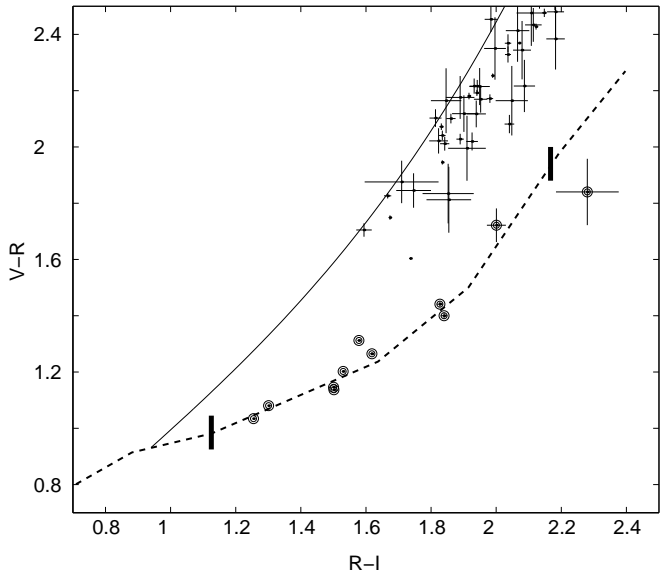}}
\caption{Star brighter than the completeness limit, including the final foreground M dwarfs in the Orion A area. Errors are 1\,$\sigma$. The dashed line is the main sequence. The solid line is a fit to the reddened background stars in $\rho$ Oph and Serpens. Double ring objects are the finally accepted sample of foreground candidates, fitting as such both by colour and photometric distance. The two reddest final candidates were also measured in $J$ and thereby confirmed to be unreddened objects}
\label{Figorionsel}
\end{figure}

\begin{figure}
\resizebox{\hsize}{!}{\includegraphics{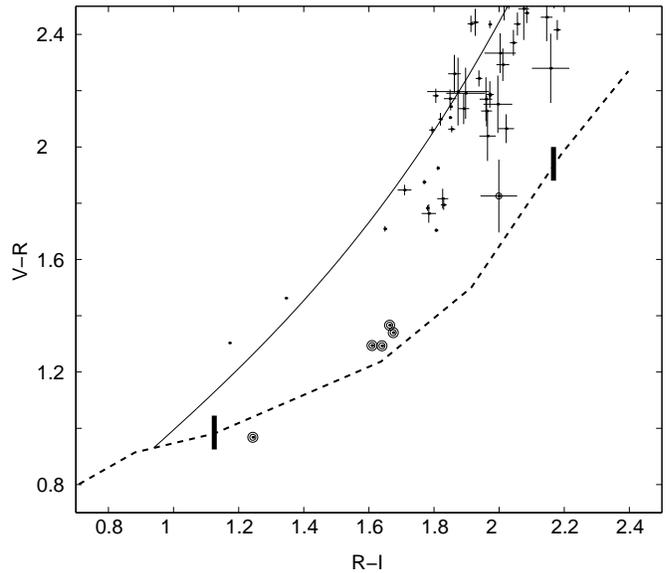}}
\caption{Star brighter than the completeness limit, including the final foreground M dwarfs in the Serpens area. Symbols are as in Fig.~\ref{Figorionsel}. Single rings are foreground candidates by colour, but not by distance}
\label{Figserpsel}
\end{figure}

\begin{figure}
\resizebox{\hsize}{!}{\includegraphics{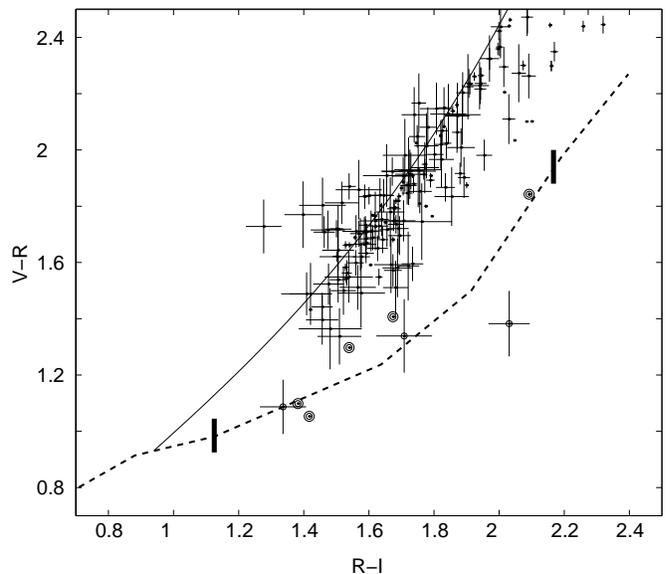}}
\caption{Star brighter than the completeness limit, including the final foreground M dwarfs in the $\rho$ Oph area. Symbols are as in Fig.~\ref{Figorionsel} \&~\ref{Figserpsel} }
\label{Figophsel}
\end{figure}

The reddened background stars separate from the foreground main sequence at $V-R=R-I=0.9$ (M0V), resulting in two parallell sequences separated by $\sim$\,0.5 mag in $V-R$ (Fig.~\ref{Figorionsel}). All stars within 0.15 mag of the empirical main sequence were considered as foreground candidates by colour and had their photometric parallaxes measured (see Sect. below). The resulting foreground M-dwarf sample is shown in Figs.~\ref{Figorionsel} -~\ref{Figophsel} together with all reddened background stars brighter than the completeness limit.

 In $\rho$ Oph a number of bright stars fall inbetween the main sequence and the reddened background stars. This is not a result of systematic photometric errors, since in the same field there are many stars lying perfectly on the reddening sequence. These objects are preferably found in $\rho$ Oph W. Since they are absent from the Serpens area at a lower galactic latitude, reddened M giants are not a likely explanation. Most probably they are recently formed stars, still embedded in the nebula (see also Sect.~\ref{crosscorr} and Fig.~\ref{Figothers}).  

\begin{figure}
\resizebox{\hsize}{!}{\includegraphics{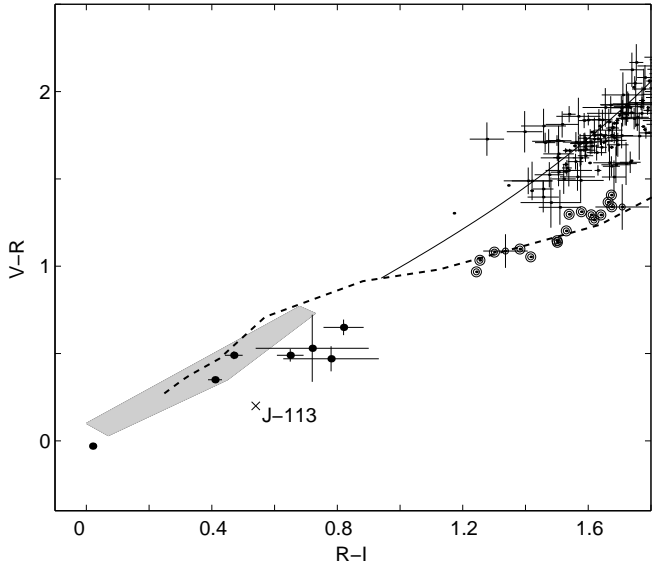}}
\caption{Photometric WD candidates. Symbols are as in Figs.~\ref{Figorionsel}-~\ref{Figophsel}. New symbols: The shaded area outlines the WDs of Bergeron et al. (\protect\cite{bergeron97}). Errorbars are 1\,$\sigma$. J-113 is a WD from JDH94}
\label{Figmdwdsel}
\end{figure}

As photometric WD candidates those objects lying approximately in the region of the WDs in Bergeron et al. (\cite{bergeron97}) were considered (Fig.~\ref{Figmdwdsel}). These objects are distinctly separated from other stars and form a well-defined isolated group. Note that the reddest WDs systematically fall below the empirical sequence. This is probably caused by the increasingly different spectra in WDs as compared to the standard stars towards cooler temperatures. 

\subsection{Photometric parallaxes}
Photometric distances, although suffering from a large spread in the $M_\mathrm{V}$ vs $V-I$ relation, are used here as an additional tool for cleaning out background stars. As a  bonus, cloud distances get independent consistency checks. Note the lack of stars between 400 and 500 pc for Orion A in Fig.~\ref{Figmdhist}, clearly indicating that $d=480$ pc is an overestimate of the distance.

To check possible foreground extinction, stars within $7\degr$ from the target areas were selected from the UBV catalogue Mermilliod (\cite{mermilliod87}) via SIMBAD. $E_{B-V}$ was determined by the Q method for stars earlier than A0 (see e.g. Mihalas \& Binney (\cite{mihalas81})). Later stars with known spectral types were also used. For all three areas it was concluded that foreground extinction is too small to affect the results of this paper.

\subsubsection{M dwarfs}
\begin{figure}
\resizebox{\hsize}{!}{\includegraphics{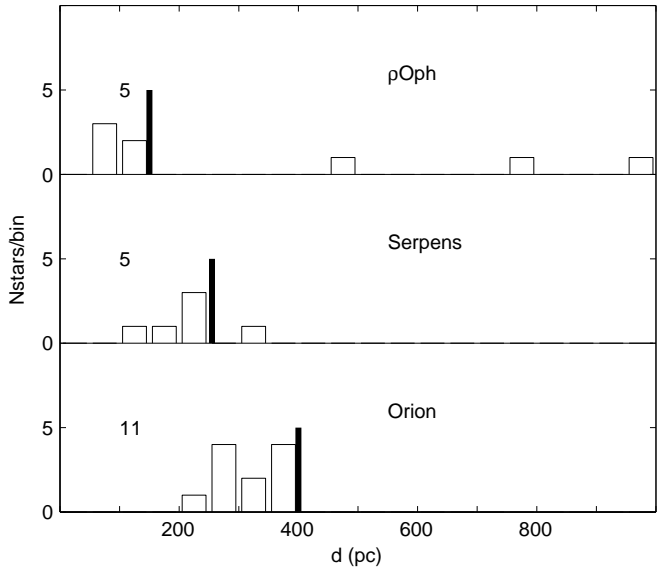}}
\caption{Photometric mean distances to the photometric foreground M-dwarf candidates. The solid bars show the adopted cloud distances}
\label{Figmdhist}
\end{figure}

 Each object was assigned a mean, min and max distance. The mean distance is derived from the composed $M_\mathrm{V}$ vs $V-I$ relation in RG97. Min and max values come from their corresponding $1\,\sigma$ spread . A histogram of the mean distances to the colour-colour foreground candidates marked in Figs.~\ref{Figorionsel}-~\ref{Figophsel} is shown in Fig.~\ref{Figmdhist}. Stars consistent with being foreground objects are distinctly sorted out. The final sample of probable foreground M dwarfs consists of 21 stars (Table~\ref{Tabfg}). 

\subsubsection{White dwarfs}
Photometric distances were estimated via a linear least-square fit of $M_\mathrm{V}$ vs $V-I$ of the WDs in Bergeron et al. (\cite{bergeron97}) with parallax errors less than 30\,\% :\\

$
\begin{array}{lllr}
\makebox[0.5cm][l]{\hfill}M_\mathrm{V} & = & 12.67 + 2.48\,(V-I) &\makebox[1.0cm][l]{\hfill} (\sigma = 0.3)
\end{array}
$
\\

This functional form provides only a rough estimate of the WD absolute magnitude. One should keep in mind that there are WDs like ESO439-26 (Ruiz et al. \cite{ruiz95b}) that do not follow this relation in the least.    

It is possible, although unlikely, that these WD candidates are reddened background stars. Since the background extinction at the position of the WD candidates in all cases is estimated to exceed 6.8 mag in $V$ (Table ~\ref{Tabfg}), a through-shining star of the same colour as our reddest WD candidate ($V-I\sim1.5$) would have an unreddened colour $(V-I)_{0}<-1.2$. There are no such stars. Suppose that they are indeed early-type stars shining through low-extinction windows. A B1V star behind OrionA would then have to be at a distance of about 0.5 Mpc to fit the observed magnitudes, i.e 160 kpc above the galactic plane, which is unrealistic. The simplest answer is that these objects are indeed foreground WDs.

\subsubsection{Spectroscopic confirmation of white dwarfs}
\begin{figure}
\resizebox{\hsize}{!}{\includegraphics{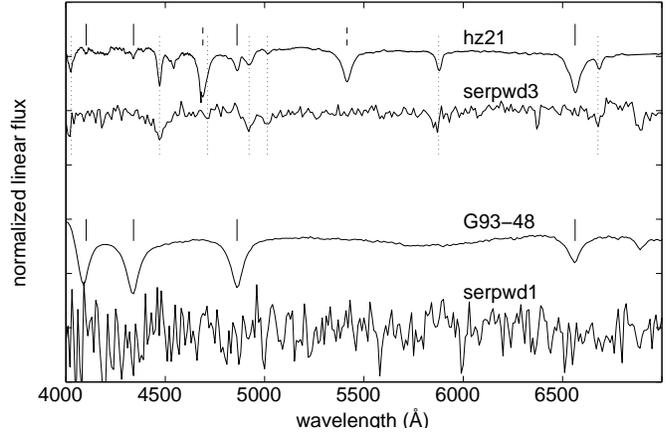}}
\caption{Low-resolution spectra of two of the Serpens WDs. The dotted lines show HeI lines, dased lines are HeII and solid lines H. Note the sky lines at $\lambda$$\lambda$5890, 6867 (Na D, and atmospheric B band)}
\label{Figwdspectra}
\end{figure}

In Fig.~\ref{Figwdspectra}, spectra taken at the NOT of the two brightest WDs are compared with two spectrophotometric standard stars. Hz21 is a He-rich WD of type DO, while G93-48 is of type DA. The serpwd3 spectrum is classified as type DB, since it only shows HeI lines (Dreizler \& Werner \cite{dreizler96}). The second WD is serpwd1, showing a rather noisy spectrum due to its faint magnitude. However, the absence of prominent spectral features in combination with its magnitude and colour makes it a very likely WD.

\begin{table*}
\caption[]{Final foreground candidates. Photometric errors are internal and 1\,$\sigma$. The photometric system is $RI$ Kron-Cousins (KC) and $V$ Johnson (J). Upper and lower distance limits are also 1\,$\sigma$ ($\sigma$ originating mainly from cosmic spread in the $M_\mathrm{V}$ - $V-I$ relation. $A_\mathrm{V}$ is the lower limit of the background extinction at each star's position. The last column is the weighted count each star gets when reduced to the solar neighbourhood}
\label{Tabfg}
\begin{flushleft}
\begin{minipage}[t]{10cm}
\begin{tabular}{lrrccccrrrrc}
\hline
 \hspace{0.2cm}id&  RA (2000) & DEC (2000)&$I_{\rm KC}$&$R_{\rm KC}-I_{\rm KC}$&$V_{\rm J}-I_{\rm KC}$&$M_\mathrm{\rm V}$&$d_{m}$&$d_{+}$&$d_{-}$&$A_\mathrm{V}$&w.ct. \\
\hline
omd 1   &  5:41:22.7 &  -8:22:21 & 15.77  0.01 & 1.50  0.01 &  2.64  0.01 & 11.6 &    230 &  195 &  270 &  6.8 & 1.32\\
omd 2   &  5:41:38.6 &  -8:30:18 & 15.95  0.01 & 1.26  0.01 &  2.29  0.01 & 10.6 &    346 &  295 &  407 &  6.8 & 1.32\\
omd 3   &  5:41:14.0 &  -8:38:19 & 15.99  0.01 & 1.30  0.01 &  2.38  0.01 & 10.8 &    324 &  276 &  381 &  6.8 & 1.32\\
omd 4   &  5:41:17.5 &  -8:29:44 & 16.10  0.01 & 1.50  0.01 &  2.65  0.01 & 11.7 &    267 &  228 &  314 &  6.8 & 1.32\\
omd 5   &  5:41:04.5 &  -8:38:23 & 17.26  0.01 & 1.53  0.01 &  2.74  0.01 & 12.2 &    381 &  308 &  471 &  6.8 & 1.32\\
omd 6   &  5:41:14.8 &  -8:25:45 & 17.34  0.01 & 1.58  0.01 &  2.89  0.01 & 13.0 &    290 &  235 &  358 &  6.8 & 1.32\\
omd 7   &  5:41:19.8 &  -8:30:48 & 17.87  0.01 & 1.62  0.01 &  2.89  0.01 & 13.0 &    370 &  300 &  458 &  6.8 & 1.32\\
omd 8   &  5:41:31.8 &  -8:22:39 & 18.09  0.01 & 1.83  0.01 &  3.27  0.01 & 14.3 &    262 &  232 &  295 &  6.8 & 1.32\\
omd 9   &  5:41:23.7 &  -8:35:42 & 18.35  0.01 & 1.81  0.01 &  3.21  0.01 & 14.2 &    300 &  266 &  338 &  6.8 & 1.32\\
omd10\footnote{$J_{CIT}=17.77$ (0.06)}   &  5:41:21.5 &  -8:23:09 & 19.63  0.02 & 2.00  0.02 &  3.72  0.06 & 15.5 &    382 &  339 &  430 &  6.8 & 1.32\\
omd11\footnote{$J_{CIT}=18.08$ (0.07)}   &  5:41:18.9 &  -8:27:39 & 20.12  0.01 & 2.33  0.04 &  4.17  0.11 & 16.6 &    358 &  317 &  403 &  6.8 & 1.32\\ [0.2cm]
serpmd1 & 18:27:10.8 &  -3:43:59 & 16.66  0.01 & 1.64  0.01 &  2.94  0.01 & 13.2 &    186 &  151 &  230 &  7.6 & 1.03\\
serpmd2 & 18:27:57.2 &  -3:31:37 & 15.02  0.01 & 1.25  0.01 &  2.21  0.01 & 10.3 &    241 &  205 &  283 & 11.3 & 1.03\\
serpmd3 & 18:27:53.8 &  -3:36:53 & 16.10  0.01 & 1.68  0.01 &  3.02  0.01 & 13.6 &    125 &  110 &  140 &  8.9 & 1.03\\
serpmd4 & 18:28:32.8 &  -3:34:53 & 16.81  0.01 & 1.61  0.01 &  2.91  0.01 & 13.1 &    212 &  171 &  262 &  9.9 & 1.03\\
serpmd5 & 18:29:20.3 &  -3:44:15 & 17.29  0.01 & 1.66  0.01 &  3.03  0.01 & 13.7 &    212 &  189 &  240 & 12.1 & 1.03\\ [0.2cm]
ophmd1  & 16:33:02.6 & -24:30:14 & 14.60  0.01 & 1.54  0.01 &  2.84  0.01 & 12.7 &     88 &   71 &  108 &  4.8 & 1.09\\
ophmd2  & 16:32:42.8 & -24:26:55 & 15.93  0.01 & 1.67  0.01 &  3.08  0.01 & 13.8 &    109 &   97 &  123 &  6.6 & 1.09\\
ophmd3  & 16:27:30.4 & -24:32:34 & 14.43  0.01 & 1.38  0.01 &  2.48  0.01 & 11.1 &    143 &  122 &  168 &  7.8 & 1.09\\
ophmd4  & 16:27:36.5 & -24:28:33 & 13.46  0.01 & 1.42  0.01 &  2.47  0.01 & 11.1 &     92 &   78 &  108 &  7.8 & 1.09\\
ophmd5  & 16:27:26.6 & -24:25:54 & 15.99  0.01 & 2.09  0.01 &  3.93  0.01 & 16.0 &     60 &   53 &   68 &  7.8 & 1.09\\ [0.4cm]
   &   & &$I_{\rm KC}$&$R_{\rm KC}$&$V_{\rm J}$&&&&&&\\
\hline
owd 1   &  5:41:37.7 &  -8:36:09 & 22.00  0.14 & 22.78  0.06 & 23.25  0.04 & 15.8 &    313 & 269 &  365 &  6.8 & 1.32\\
owd 2   &  5:41:04.0 &  -8:23:09 & 21.24  0.03 & 21.89  0.03 & 22.38  0.02 & 15.5 &    238 & 207 &  273 &  6.8 & 1.32\\
owd 3   &  5:41:22.6 &  -8:22:50 & 22.83  0.10 & 23.55  0.15 & 24.08  0.12 & 15.8 &    459 & 393 &  537 &  6.8 & 1.32\\ [0.2cm]
serpwd1 & 18:28:05.6 &  -3:40:02 & 20.38  0.02 & 20.79  0.01 & 21.14  0.01 & 14.6 &    208 & 181 &  239 &  9.6 & 1.03\\ 
serpwd2 & 18:27:45.0 &  -3:46:01 & 20.51  0.02 & 20.98  0.02 & 21.47  0.01 & 15.0 &    192 & 167 &  221 &  9.9 & 1.03\\
serpwd3 & 18:27:41.6 &  -3:40:38 & 17.88  0.01 & 17.90  0.01 & 17.87  0.01 & 12.6 &    111 &  96 &  127 & 12.6 & 1.03\\
serpwd4 & 18:28:26.3 &  -3:43:52 & 21.14  0.05 & 21.96  0.04 & 22.61  0.02 & 16.3 &    181 & 157 &  208 & 11.6 & 1.03\\
\end{tabular}
\end{minipage}
\end{flushleft}
\end{table*}

\section{Constraints on foreground candidates from other surveys}
\label{crosscorr}
The Orion A area (L1641) has been extensively surveyed in the infrared (Strom et al. (\cite{strom89}, \cite{strom93}); Ali \& Depoy \cite{ali95}). In Strom et al. (\cite{strom89}) there is one overlapping object, however not a foreground one. The other two surveys are deeper but unfortunately do not overlap.

\begin{figure}
\resizebox{\hsize}{!}{\includegraphics{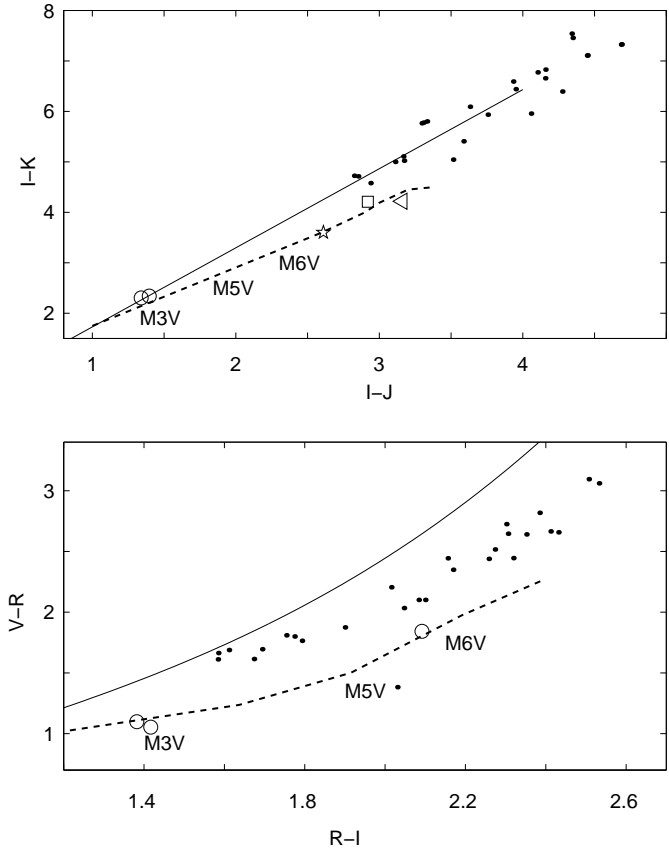}}
\caption{Overlapping near-IR sources. The two earliest foreground candidates are represented in the upper panel by mean values of the $JK$ magnitudes given in Barsony et al. (\protect\cite{barsony97}), Strom et al. (\protect\cite{strom95}) and Greene et al. (\protect\cite{greene92}). Individual values were used for the latest candidate, triangle ( Barsony), pentagram (Strom) and box (Greene).  Dashed lines are the corresponding main sequences and the solid lines show the empirical reddening sequences}
\label{Figothers}
\end{figure}

In $\rho$ Oph, the western area was entirely covered in $JHK$ in a survey by Barsony et al. (\cite{barsony97}). There are 30 overlapping sources brighter than our completeness limits, including all three foreground candidates in this area (ophmd3--5 in Table~\ref{Tabfg}). In Fig.~\ref{Figothers} the appearance of the overlapping sample in $VRI$ is compared to $IJK$. 
Ophmd3 and 4 are confirmed as foreground by the additional $JHK$ data. For ophmd5, the three independent $JHK$ measurements are not consistent, implying either variability or a problem in the photometry. The fact that ophmd5 does not keep its position relative to the bulk of the background stars between $VRI$ and $IJK$ supports it being a foreground star of spectral type about M6V.

The $\rho$ Oph E area was covered in $VRI$ by Jarrett (JDH94 + thesis).  
Both our candidates in this area (ophmd1--2) have magnitudes consistent with theirs, and are considered as possible foreground stars in Jarrett (\cite{jarrett92}). Ophmd2, though, was rejected from the foreground candidates in JDH94. Both stars are slightly redder than the main sequence in $V-R$ (Fig.~\ref{Figophsel}), but are still regarded as foreground stars. See Sect.~\ref{sec-jarrett} for a thorough comparison to Jarrett's results.

In our Serpens field, no complementary photometric data could be found.

\section{The luminosity function}
A simplified version of the $V_\mathrm{max}$ method (Felten \cite{felten76}), outlined below, was used to derive the LF. In a photometric survey each star has a maximum distance, $d_\mathrm{max}$, at which it still can be detected. The corresponding maximum volume, $V_\mathrm{max}$, is defined by the survey area and $d_\mathrm{max}$. Each star's contribution to the LF is then $V_\mathrm{max}^{-1}$. In order to compensate for galactic density gradients and reduce the derived LF to the solar neighbourhood, $V_\mathrm{max}$ must be replaced by $V_\mathrm{max,gen}$, a generalized volume (Tinney \cite{tinney93a}):\\

$
\begin{array}{lllr}
\makebox[0.5cm][l]{\hfill}V_\mathrm{max,gen} & = & \Omega \int_{0}^{d_\mathrm{max}} r^{2} e^{-rsin|b|/h} dr, &\makebox[1.5cm][l]{\hfill}
\end{array}
$
\\\\
where $\Omega$ denotes the angular area of the field, $h$ the disk scale height and $b$ the galactic latitude. The distances involved here are small enough to neglect the radial density gradient in the Galaxy. Both the M-dwarf and WD distribution was approximated by the old disk population, with a scale height $h=350$ pc (Bahcall \& Soneira \cite{bahcall80}).
Deriving $V_\mathrm{max}$ is a trivial operation in a dark cloud survey, since for all stars brighter than the completeness limit (in absolute magnitude) $d_\mathrm{max}=d_\mathrm{cloud}$. This is an important point of this paper, the survey volume not being derived through a colour-magnitude relation having a large cosmic spread. Table~\ref{Tabvol} gives the physical and generalized values of the foreground volumes. The last column in Table ~\ref{Tabfg} shows the weighted count for each foreground star deduced as $V_\mathrm{max}/V_\mathrm{max,gen}$. Errors in the distances to the clouds are expected to be on the order of 10\,\%, implying volume uncertainies of $\sim$ 30\,\%. This uncertinty has not been included in the further analysis. 

\begin{table}
\caption[]{Observed volumes in pc$^{3}$}
\label{Tabvol}
\begin{flushleft}
\begin{tabular}{lrr}
\hline
 area&$V_\mathrm{max}$&$V_\mathrm{max,gen}$ \\
\hline
 Orion A       & 285 & 216  \\ 
 Serpens       & 193 & 188  \\  
 $\rho$ Oph    &  66 &  60  \\ [0.3cm]
total          & 544 & 464  \\
\hline
\end{tabular}
\end{flushleft}
\end{table}

\subsection{M dwarfs, main sequence}
\begin{figure}
\resizebox{\hsize}{!}{\includegraphics{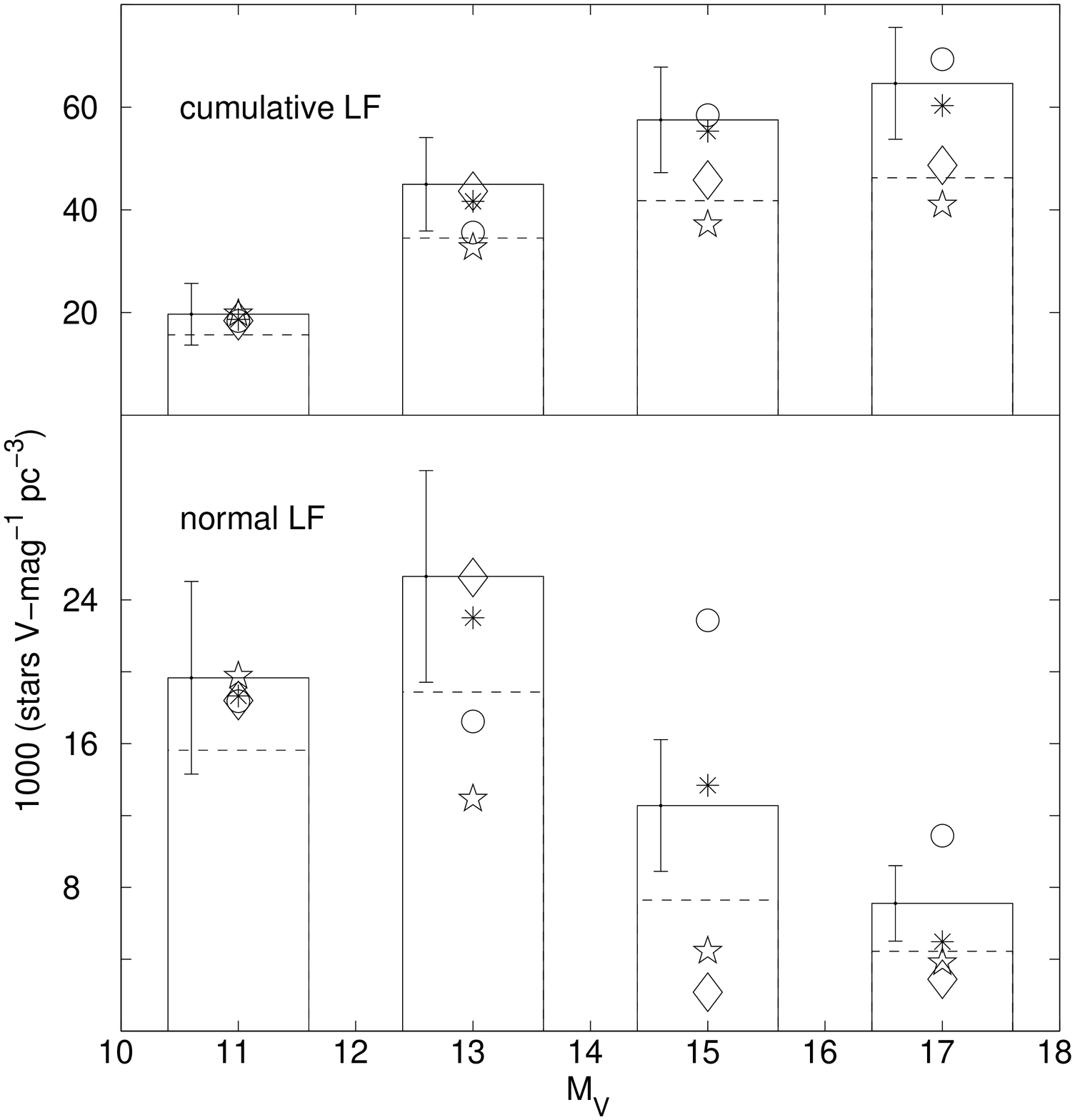}}
\caption{The M-dwarf luminosity function. The dashed line is the system LF and the solid line the single-star LF derived after binary correction. Diamonds and stars show the LF of Kirkpatrick et al. (\protect\cite{kirkpatrick94}) and Gould et al.  (\protect\cite{gould96}) respectively. Asterisks and circles outline the LF of the nearby stars in RG97 and Jahreiss \& Wielen (\protect\cite{jahreiss97}) respectively. The error bars show Poisson errors of the generalized counts before binary correction. Errors in the LF arising from uncertain cloud distances have not been included (expected to be $\sim$ 30\,\%). The incompleteness in the last bin (see Table ~\ref{TabLF}) has not been corrected for}
\label{FigLF}
\end{figure}

The majority of foreground stars are early-type M dwarfs, $2.5<V-I<3.2$, reflecting the well-known peak in the LF. The low number of M dwarfs (21) implied a rather coarse binning in the LF. Table ~\ref{TabLF} gives the LF in 1-mag bins, while in Fig.~\ref{FigLF} 2-mag bins were chosen. The agreement with previously determined LFs is satisfactory. 

For stars brighter than $M_\mathrm{V}=10$ ($\sim$ M1V) the number that is lost due to saturation starts to become significant. These stars are therefore not considered further.

The large spread in the faint bins in Fig.~\ref{FigLF} illustrates the present uncertainty in the stellar LF. The Poissonian 1 $\sigma$ uncertainties in Jahreiss \& Wielen (\cite{jahreiss97}) and RG97 are about 30\,\% in the last two bins. 

\begin{table}
  \caption[]{The foreground M-dwarf sample binned into 1-mag bins in $M_\mathrm{V}$. The 2nd last bin is complete in 70\,\% of the volume and the last bin in 26\,\%. No correction for incompleteness has been applied}
  \label{TabLF}
  \begin{flushleft}
    \begin{tabular}{lllllll}
      \hline
       bin & \multicolumn{3}{c}{systems} & &\multicolumn{2}{c}{binary corrected} \\ 
      \cline{2-4}\cline{6-7}
      $M_\mathrm{V}$& $N$&$N_\mathrm{gen}$& cum          & &$N_\mathrm{gen}$& cum                \\ 
      \hline
       10-11       &   3 & 3.67 &     3.67 &  & 4.49 &  4.49\\  
       11-12       &   4 & 4.82 &     8.49 &  & 6.19 & 10.68\\	
       12-13       &   2 & 2.41 &    10.90 &  & 4.46 & 15.14\\	
       13-14       &   7 & 7.85 &    18.75 &  & 9.30 & 24.45\\	
       14-15       &   2 & 2.64 &    21.39 &  & 4.20 & 28.65\\	
       15-16       &   1 & 1.32 &    22.71 &  & 2.62 & 31.27\\	
       16-17       &   2 & 2.41 &    25.12 &  & 3.06 & 34.33\\	
       17-18       &   0 & 0    &    25.12 &  & 0.80 & 35.13\\ 
      \hline
    \end{tabular}
  \end{flushleft}
\end{table}

\subsubsection{Lost companions}
RG97 found that the number of lost unresolved companions approximately is compensated for by distant stars spread into a magnitude limited sample via Malmquist bias. This will not be the case in this survey, since the foreground sample is strictly volume limited and there is a dark cloud preventing distant stars from interfering. The number of lost companions was estimated from the binary distribution in the nearby stars in RG97. After applying the correction for galactic structure 10 companions were expected to have been lost. These were distributed as the lost binary companions in Fig. 12 of RG97 and added to our system LF.
The resulting companion-corrected LF is shown in  Fig.~\ref{FigLF} and Table~\ref{TabLF}.

\subsection{White dwarfs}
The sample of WDs is too small to derive any details in the WD LF. Seven WDs were found, 0.015 ($\pm 0.006$)\,pc$^{-3}$ (Poissonian errors), which is a significantly higher number than expected from LDM88 (0.003\,pc$^{-3}$), but in better agreement with RT95 and Oswalt et al. (\cite{oswalt96}), who both give $\sim\,0.008$\,pc$^{-3}$. These numbers will be dicussed in more detail in Sect.~\ref{wddiscuss}.

\section{Dicussion, comparison to other works}
For the M-dwarf part, the agreement with other photometric surveys is satisfactory (Fig.~\ref{FigLF}). It is clear that any contamination of the M-dwarf sample from background stars and/or stars associated with the clouds must be small and that the opaque-nebula method works reliably also in a purely photometric version, provided that the photometry is sufficiently accurate.

The main advantage in comparison to other methods is that the survey volume does not rely on photometric parallaxes. Even though the individual absolute magnitudes do suffer from cosmic spread in the colour-magnitude relation, each star's contribution to the space density does not. 

If the distance to the cloud is not known a priori, it can either be estimated from the photometric distances of the foreground objects or by their integrated number in a well-known part of the LF. In this context it would be useful to obtain more data for foreground stars brightwards of the peak in the LF at $M_\mathrm{V}=12$. 

The derived samples are volume limited and unaffected by dynamical bias. Finally, the dark clouds effectively screen background galaxies and giant stars from the foreground population.
  
Among the drawbacks are that the selected foreground objects in general will be too distant for binaries to be resolved and therefore the derived LF must be corrected for by an a priori known binary distribution function. The dark clouds are not evenly spread across the sky, which means that the available areas are restricted. On the other hand, they do provide a tool for probing areas near the Milky Way plane.  

\subsection{The Jarrett paper}
\label{sec-jarrett}

\begin{figure}
\resizebox{\hsize}{!}{\includegraphics{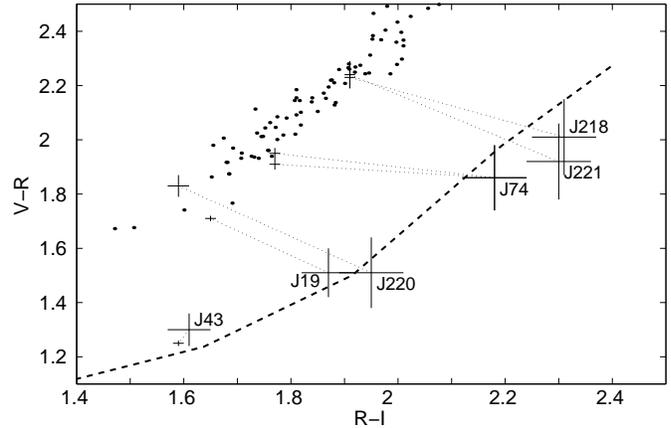}}
\caption{A comparison with JDH94. The dashed line marks the main sequence and the dots outline the reddening sequence. The + signs mark the overlapping objects with 1\,$\sigma$ errorbars. The JDH94 counterparts are marked with Jx, x indicating the id in Table 3 in JDH94. The dotted lines show the deduced colour differences. For J74 there are two independent measurements in our survey}
\label{Figjarrett}
\end{figure}

We compare our result with that of JDH94. The generalized volume sampled in that paper is 52 pc$^{3}$ ($\rho$ Oph) + 101 pc$^{3}$ (Taurus), totally 153 pc$^{3}$. A striking difference as compared to our survey is their excess of late foreground candidates, seen also in the Barnard\,5 survey (Jarrett \cite{jarrett95}). The comparison here was made in $\rho$ Oph E, where part of our surveys overlap. Five of their six foreground candidates that were cross identified lie perfectly on the reddening sequence in our data and are definitely not foreground objects (Fig.~\ref{Figjarrett}). Since our data for the bright J43 are consistent, the photometric calibration is likely to be correct. A possible explanation of the colour discrepancy is that these sources were spread into the JDH94 foreground sample as a result of their lower signal-to-noise ratio. Our data reach $\sim$ 2 magnitudes deeper than JDH94. It may also be that the JDH94 photometry is subject to systematically discrepant corrections for atmospheric correction, since their data were acquired at high air masses ($\sim2$, Jarrett, priv comm.). The tendency that very faint stars, barely measurable, scatter into the M-dwarf zone is seen also in our data, and was the main reason for including objects only to the completeness limit and not all the way down to the detection limit of the survey, about 1 mag fainter. 

\subsection{White dwarfs}
\label{wddiscuss}
A local number density of 0.015 ($\pm0.006$) WDs\,pc$^{-3}$ was derived from the 7 WDs discovered in this survey. This is a factor 5 more than predicted from LDM88. RT95 found 0.008 ($\pm 0.003$) WDs\,pc$^{-3}$ in a proper motion survey, consistent with our result, but as in our case the numbers are small. 
An interesting result is that Ruiz et al. (\cite{ruiz93}) found a factor 2 more stars with $\mu>0.5"$\,yr$^{-1}$ than present in \emph{LHS} (Luyten \cite{luyten76}) for overlapping areas. The LDM88 sample was drawn from \emph{LHS} and would also be affected by this proposed incompleteness, a fact that may explain the discrepancy in the LDM88 WD density as compared to more recent results.

Oswalt et al. (\cite{oswalt96}) used yet another approach and arrived at $0.0053^{+.0035}_{-0.0007}$ WDs\,pc$^{-3}$ for WDs in wide binaries. They added the single-WD LF in LDM88 (0.0023 WDs\,pc$^{-3}$) and concluded that the total space density of WDs in the solar neighbourhood is $0.0076^{+.0037}_{-0.0007}$ WDs\,pc$^{-3}$. Our results and RT95 show that the LDM88 sample is likely to suffer from a rather large incompleteness, suggesting that adding 0.0023 WDs\,pc$^{-3}$ to the binary part results in a substantial underestimation of the total WD space density. 

Our sample of WDs includes neither wide nor narrow binaries, except possible double WDs. If we add our WD density to the Oswalt binaries the result is 0.02 ($\pm0.007$) WDs\,pc$^{-3}$, corresponding to 0.013 ($\pm0.005$)\,M$_{\sun}$\,pc$^{-3}$ if we assume a mean mass of 0.65\,M$_{\sun}$ per WD. The WD fraction of the dynamical mass would then be 17 ($\pm7$)\,\%. Objects like ESO 439-20 (RT95) and WD 0346+246 (Hambly et al. \cite{hambly97}) suggest that the WD density may be even larger.

The results of the MACHO micro lensing experiment (Alcock et al. \cite{alcock97}) have given a mean mass of the lensing objects of $\sim$\,0.5 M$_{\sun}$, although strongly model dependent. The MACHO mass density in the solar neighbourhood is on the order of 0.005 M$_{\sun}$\,pc$^{-3}$, which in fact is consistent with the proposed increase of the WD space density. However, more definite results will have to await a larger volume limited sample of WDs.

\section{Conclusions and future prospects}
The space density of M dwarfs and WDs has been studied as volume limited samples in front of dark high extinction nebulae, a method that efficiently reduces the impact of Malmquist bias, since the survey is volume limited and the LF not dependening on individual photometric distance estimates. Areas towards the molecular cloud complexes $\rho$ Oph, Serpens and Orion A were observed. The survey covers a volume corresponding to 464 pc$^{3}$ in the solar neighbourhood, complete to $M_\mathrm{V}\sim16.5$. In total 21 M dwarfs and 7 WDs were identified as foreground objects.

Although details are hard to resolve due to the limited sample, our M-dwarf LF is consistent with previous photometric results, and after binary correction also with the nearby stars. 
Our survey does not show any signs of the reported upturn in the faintest magnitude bins in the previous dark cloud surveys (Jarrett \cite{jarrett92},\cite{jarrett95}; Jarrett et al. \cite{jarrett94}),  and we conclude that their excess of faint M dwarfs is probably caused by their larger photometric errors. 

The unexpected appearance of as many as 7 WDs compared to the 1.4 estimated from the WD space density in LDM88 is perhaps the most interesting result of this paper. The derived space density is 0.015 ($\pm0.006$) WDs\,pc$^{-3}$. By combining our sample with the binary WDs in Oswalt et al. (\cite{oswalt96}) the final estimate is 0.02 ($\pm0.007$) WDs\,pc$^{-3}$, implying a mass density of 0.013 ($\pm0.005$) M$_{\sun}$\,pc$^{-3}$. This means that the WDs may contribute by as much as 15--20\,\% to the local dynamical mass. 
No doubt this subject is not yet settled and deserves further investigation.\\

The method of opaque nebulae has indeed proved to be efficient in identifying foreground M dwarfs and WDs. Concerning the latter, any nebula that provides a colour excess $E_{V-I}>2$ ($A_\mathrm{V}>5$) ensures that also the bluest main sequence stars get a reddened colour beyond the reddest known WDs ($V-I\sim1.5$). Already for Orion A the available area is $\sim 50$ times larger than our field, enclosing a volume that would host $\sim$\,150 WDs and $\sim$\,550 M dwarfs, numbers that certainly are sufficient for a more detailed analysis of the corresponding LFs.
The increased availability of large field CCDs and CCD mosaics provides the ideal tools for this kind of survey.
We have also shown that the distance to the clouds and the cloud extinction may be reliably derived from the $VRI$ data alone, an important point if a future large-scale survey of this kind is to be carried out.

\begin{acknowledgements}
      I wish to thank the ESO OPC for allocating observing time for this exciting project.
      This work was supported by the Nordic Optical Telescope Scientific 	
      Association and the Swedish Natural Science Research Council (NFR). This research has made use of the Simbad database, operated at CDS, Strasbourg, France
\end{acknowledgements}

\appendix
\section{The observed areas}
\label{appA}

\begin{figure}
\resizebox{\hsize}{!}{\includegraphics{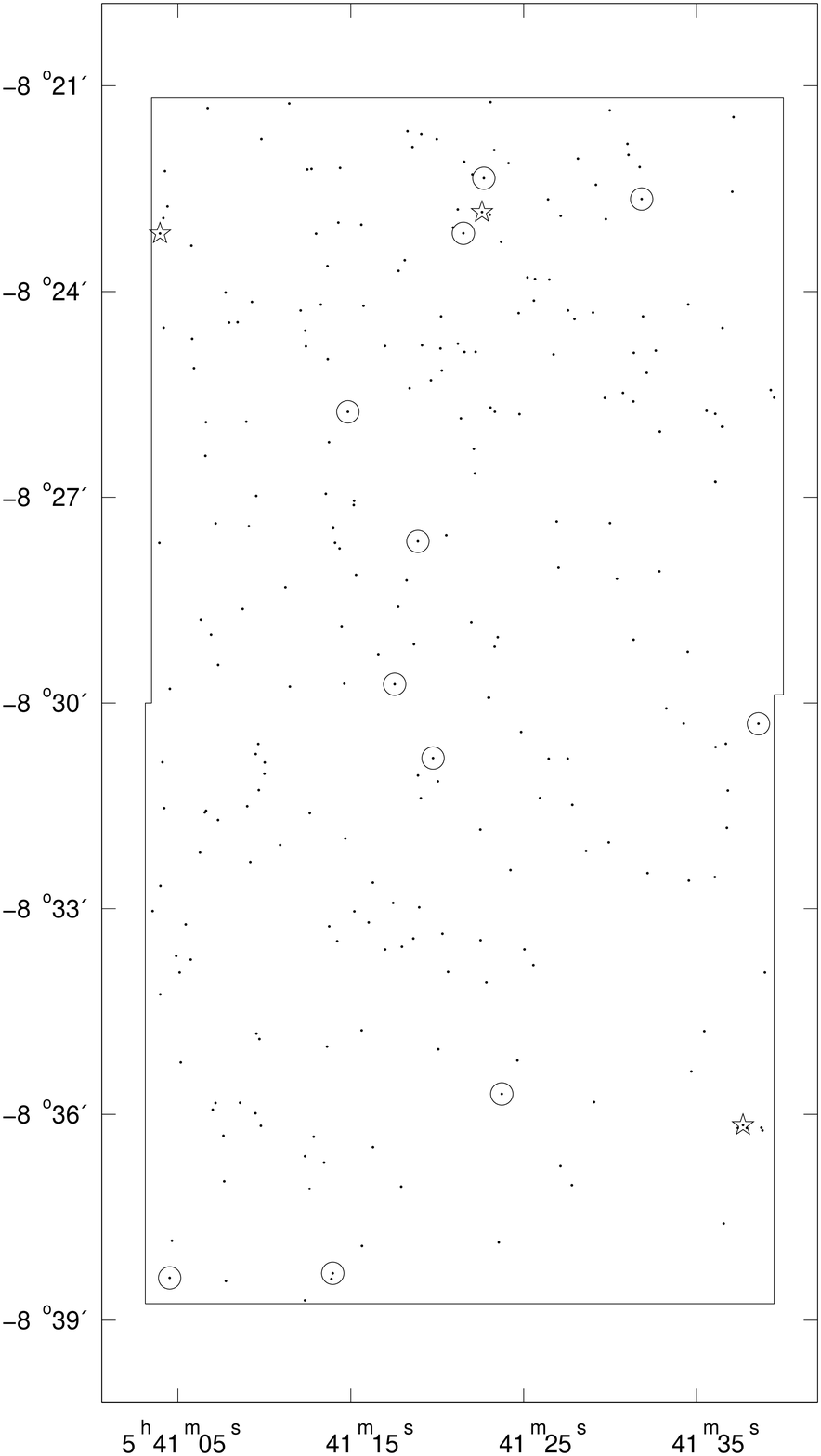}}
\caption{The Orion A area (southern part of L1641). Coordinates are equinox 2000. Small dots mark all objects observed in $I$. Rings are the foreground M dwarfs. Pentagrams are white dwarfs}
\label{Figstarsorion}
\end{figure}

\begin{figure*}
\resizebox{\hsize}{!}{\includegraphics{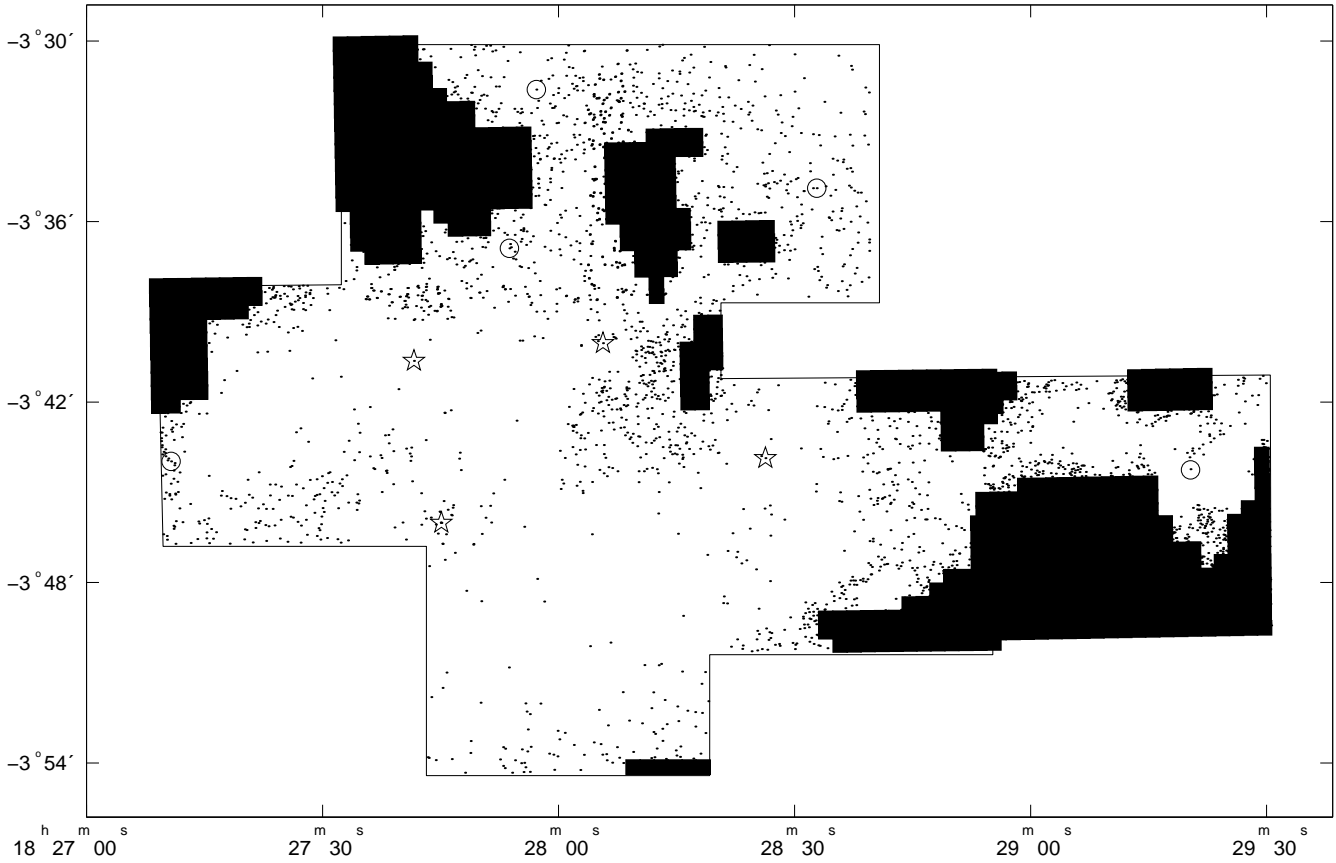}}
\hfill
\caption{Serpens ($\sim 4\degr$ south of the Serpens cloud core). Coordinates are equinox 2000. Symbols as in Fig. ~\ref{Figstarsorion}. Filled areas were rejected from the survey due to high stellar density (Sect. ~\ref{secextract})}
\label{Figstarsserp}
\end{figure*}

\begin{figure*}
\resizebox{\hsize}{!}{\includegraphics{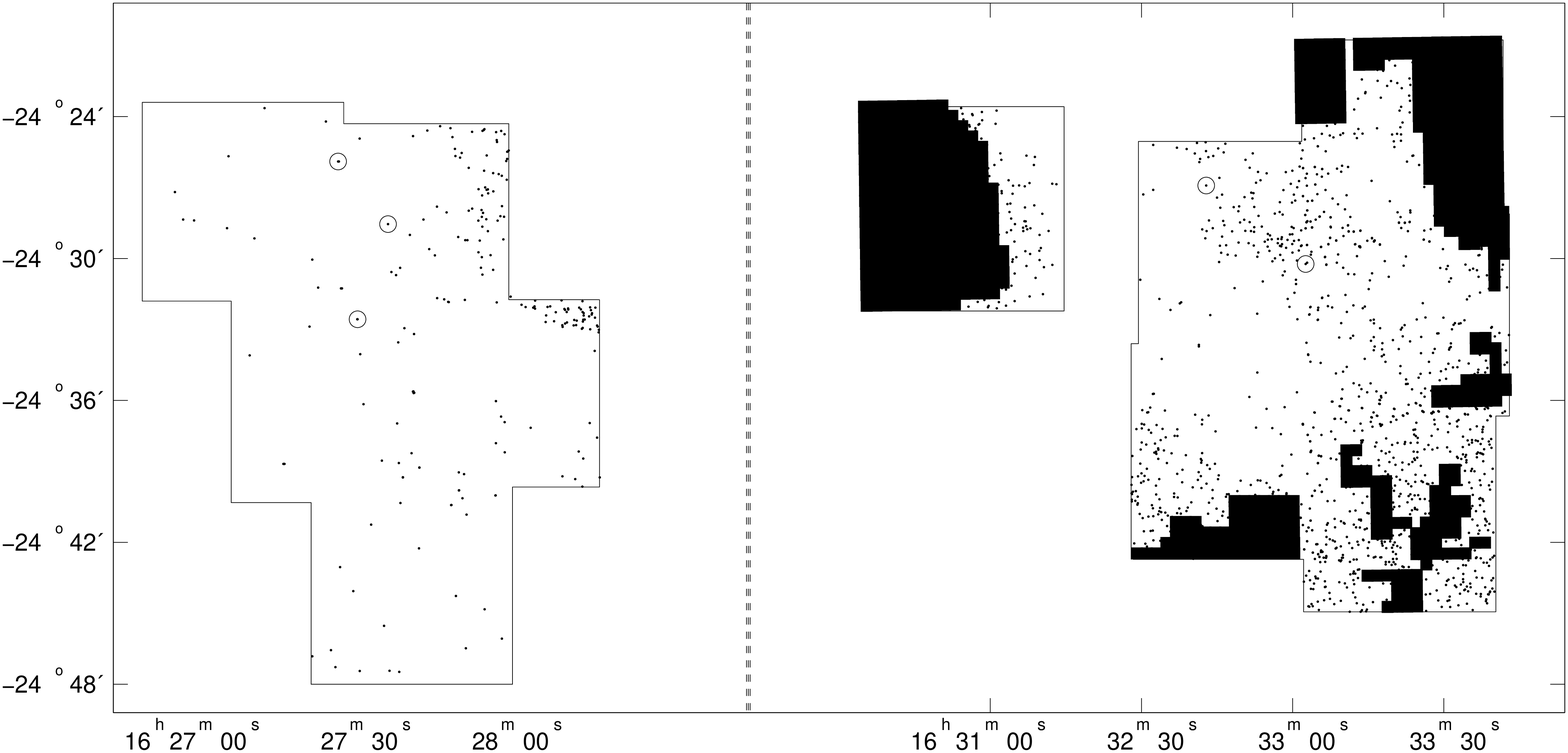}}
\hfill
\caption{$\rho$ Oph W (left, L1688) and $\rho$ Oph E (right, L1689). Coordinates are equinox 2000. Symbols as in Figs. ~\ref{Figstarsorion} - ~\ref{Figstarsserp}}
\end{figure*}


\begin{thebibliography}{}

\bibitem[1997]{alcock97}
{Alcock} C., {Allsman} R.~A., {Alves} D., et al. 1997,
\newblock ApJ 486, 697

\bibitem[1995]{ali95}
{Ali} B., {DePoy} D.~L. 1995,
\newblock AJ 109, 709

\bibitem[1982]{anthonytwarog82}
{Anthony-Twarog} B.~J. 1982,
\newblock AJ 87, 1213

\bibitem[1980]{bahcall80}
{Bahcall} J.~N., {Soneira} R.~M. 1980,
\newblock ApJS 44, 73

\bibitem[1992]{bahcall92}
{Bahcall} J.~N., {Flynn} C., {Gould} A. 1992,
\newblock ApJ 389, 234

\bibitem[1991]{bally91}
{Bally} J., {Langer} W.~D., {Liu} W. 1991,
\newblock ApJ 383, 645

\bibitem[1997]{barsony97}
{Barsony} M., {Kenyon} S.~J., {Lada} E.~A., {Teuben} P.~J. 1997,
\newblock ApJS 112, 109

\bibitem[1997]{bergeron97}
{Bergeron} P., {Ruiz} M.~T., {Leggett} S.~K. 1997,
\newblock ApJS 108, 339

\bibitem[1997]{creze97}
{Cr\'ez\'e} M., {Chereul} E., {Bienayme} O., {Pichon} C. 1997,
\newblock in: B. {Battrick} (ed.), Hipparcos Venice '97, p. 669, ESA SP-402

\bibitem[1978]{dickman78}
{Dickman} R.~L. 1978,
\newblock AJ 83, 363

\bibitem[1996]{dreizler96}
{Dreizler} S., {Werner} K. 1996,
\newblock A\&A 314, 217

\bibitem[1978]{elias78}
{Elias} J.~H. 1978,
\newblock ApJ 224, 857

\bibitem[1997]{hipparcos}
{ESA} 1997,
\newblock The Hipparcos and Tycho Catalogues,
\newblock ESA SP-1200

\bibitem[1976]{felten76}
{Felten} J.E. 1976,
\newblock ApJ 207, 700

\bibitem[1997]{festin97a}
{Festin} L. 1997,
\newblock A\&A 322, 455

\bibitem[1998]{festin98a}
{Festin} L. 1998,
\newblock A\&A 333, 497

\bibitem[1991]{fukui91}
{Fukui} Y., {Mizuno} A. 1991,
\newblock in: E. {Falgarone}, F. {Boulanger}, G. {Duvert} (eds.), Fragmentation of Molecular Clouds and Star Formation: IAU Symp. 147, p. 275

\bibitem[1996]{gould96}
{Gould} A., {Bahcall} J.~N., {Flynn} C. 1996,
\newblock ApJ 465, 759

\bibitem[1992]{greene92}
{Greene} T.~P., {Young} E.~T. 1992,
\newblock ApJ 395, 516

\bibitem[1997]{hambly97}
{Hambly} N.~C., {Smartt} S.~J., {Hodgkin} S.~T. 1997,
\newblock ApJ 489, L157

\bibitem[1997]{haywood97}
{Haywood} M., {Robin} A.~C., {Cr\'ez\'e} M. 1997,
\newblock A\&A 320, 440

\bibitem[1983]{herbst83}
{Herbst} W., {Dickman} R.~L. 1983,
\newblock in: A.~G. Davis Philip, A.~R. Upgren (eds.), The Nearby Stars and the
  Stellar Luminosity Function: IAU Coll. 76, p. 187

\bibitem[1995]{hilton95}
{Hilton} J. \& {Lahulla} J.~F. 1995,
\newblock A\&AS 113, 325

\bibitem[1995]{hunt95}
{Hunt} L.~K., {Migliorini} S., {Testi} L., et al. 1995,
\newblock AJ,
\newblock submitted

\bibitem[1997]{jahreiss97}
{Jahreiss} H., {Wielen} R. 1997,
\newblock in: B. {Battrick} (ed.), Hipparcos Venice '97, p. 675, ESA SP-402

\bibitem[1992]{jarrett92}
{Jarrett} T.~H. 1992,
\newblock Ph.D. thesis, Massachusetts Univ., Amherst.

\bibitem[1994]{jarrett94}
{Jarrett} T.~H., {Dickman} R.~L., {Herbst} W. 1994,
\newblock ApJ 424, 852

\bibitem[1995]{jarrett95}
{Jarrett} T.~H. 1995,
\newblock in: C.~G. Tinney (ed.), The bottom of the main sequence -and beyond, Springer, Berlin, p. 187

\bibitem[1994]{kirkpatrick94}
{Kirkpatrick} J.~D., {McGraw} J.~T., {Hess} T.~R., {Liebert} J., {McCarthy Jr.} D.~W. 1994,
\newblock ApJS 94, 749

\bibitem[1995]{kroupa95}
{Kroupa} P. 1995,
\newblock ApJ 453, 358

\bibitem[1993]{kroupa93}
{Kroupa} P., {Tout} C., {Gilmore} G. 1993,
\newblock MNRAS 262, 545

\bibitem[1992]{landolt92}
{Landolt} A.~U. 1992,
\newblock AJ 104, 340

\bibitem[1988]{liebert88}
{Liebert} J., {Dahn} C.~C., {Monet} D.~G. 1988,
\newblock ApJ 332, 891

\bibitem[1989]{loren89}
{Loren} R.~B. 1989,
\newblock ApJ 338, 902
 
\bibitem[1976]{luyten76}
{Luyten} W.~J. 1976,
\newblock The LHS Catalogue, University of Minnesota

\bibitem[1996]{mera96}
{M\'era} D., {Chabrier} G., {Baraffe} I. 1996,
\newblock ApJ 459, L87

\bibitem[1987]{mermilliod87}
{Mermilliod} J.~C. 1987,
\newblock A\&AS 71, 119

\bibitem[1981]{mihalas81}
{Mihalas} D. \& {Binney} J. 1981,
\newblock in: Galactic astronomy: Structure and kinematics /2nd edition/, W. H. Freeman and Co., San Francisco 

\bibitem[1996]{oswalt96}
{Oswalt} T.~D., {Smith} J.~A., {Wood} M.~A., {Hintzen} P. 1996,
\newblock Nat 382, 692

\bibitem[1997]{reid97}
{Reid} I.~N., {Gizis} J.~E. 1997,
\newblock AJ 113, 2246

\bibitem[1995]{ruiz95a}
{Ruiz} M.~T., {Takamiya} M.~Y. 1995,
\newblock AJ 109, 2817

\bibitem[1995]{ruiz95b}
{Ruiz} M.~T., {Bergeron} P., {Leggett} S.~K., {Anguita} C. 1995,
\newblock ApJ 455, L159

\bibitem[1993]{ruiz93}
{Ruiz} M.~T., {Takamiya} M.~Y., {M\'endez} R., {Maza} J., {Wishnjewsky} M. 1993,
\newblock AJ 106, 2575

\bibitem[1986]{scalo86}
{Scalo} J.~M. 1986,
\newblock Fund. Cosm. Phys. 11, 1

\bibitem[1996]{straizys96}
{Straizys} V., {Cernis} K., {Bartasiute} S. 1996,
\newblock Baltic Astronomy 5, 125

\bibitem[1995]{strom95}
{Strom} K.~M., {Kepner} J., {Strom} S.~E. 1995,
\newblock ApJ 438, 813

\bibitem[1989]{strom89}
{Strom} K.~M., {Newton} G., {Strom} S.~E., et al. 1989,
\newblock ApJS 71, 183

\bibitem[1993]{strom93}
{Strom} K.~M., {Strom} S.~E., {Merrill} K.~M. 1993,
\newblock ApJ 412, 233

\bibitem[1993]{tinney93a}
{Tinney} C.~G. 1993,
\newblock ApJ 414, 254

\bibitem[1978]{warren78}
{Warren} W.~H., {Hesser} J.~E. 1978,
\newblock ApJS 36, 497

\bibitem[1997]{winkler97}
{Winkler}, H., 1997,
\newblock MNRAS 287, 481

\end{thebibliography}
\end{document}